\documentclass[12pt]{article}
\usepackage{amsmath,amssymb,amsthm,mathrsfs,url}
\usepackage{hyperref,graphicx,color,natbib}

\title{A Vision for a Bohm-Style Theory of Quantum Electrodynamics}

\author{
Roderich Tumulka\footnote{Fachbereich Mathematik, 
	Eberhard-Karls-Universit\"at T\"ubingen, 
	Auf der Morgenstelle 10, 72076 T\"ubingen, Germany. 
	E-mail: roderich.tumulka@uni-tuebingen.de}
}

\date{October 12, 2023}

\newcommand{\Hilbert}{\mathscr{H}}
\newcommand{\Fock}{\mathscr{F}}
\newcommand{\be}{\begin{equation}}
\newcommand{\ee}{\end{equation}}
\newcommand{\scp}[2]{\langle #1|#2\rangle}

\newcommand{\RRR}{\mathbb{R}}

\newcommand{\CCC}{\mathbb{C}}
\newcommand{\ZZZ}{\mathbb{Z}}

\newcommand{\vx}{\boldsymbol{x}}

\newcommand{\sM}{\mathscr{M}}

\DeclareMathOperator{\tr}{tr}

\begin{document}
\maketitle
\begin{abstract}
Despite many successes of quantum electrodynamics (QED), we do not presently have a good understanding of this field of physics. QED has all of the foundational problems that standard non-relativistic quantum mechanics has, and further ones in addition. I discuss some of these problems and some options for what a Bohm-style theory of QED, with an ontology in space and time, could look like. I also point out why the proposal made by Bohm himself in 1952 for QED is not quite convincing. Finally, I outline the kind of Bohm-type theory of QED that I would consider convincing, and report about recent progress toward this kind of theory.

\medskip

\end{abstract}

\section{Introduction}

In this paper, I would like to outline why quantum electrodynamics (QED) in its usual formulation is not very satisfactory, and what a satisfactory theory could look like. It is one of the goals of physics to find the fundamental laws of nature; I tend to take seriously the possibility that the correct laws could have a Bohmian character, meaning that there is an ontology in space and time that evolves according to laws that involve the wave function, and here I explore and discuss this possibility.

In non-relativistic quantum mechanics (NRQM), the orthodox formulation of the theory has several problems that have been discussed in many places \citep[e.g.,][]{Tum22}, in particular: that the theory does not clearly say what happens in reality; that it fails to apply the same laws to a measurement apparatus as to every other system of quarks and electrons, or else runs into inconsistencies known as the quantum measurement problem; that it remains unclear how the non-locality required by Bell's theorem \citep{Bell,GNTZ} gets reconciled with relativity; that it remains unclear where in the fundamental laws the randomness comes in; and that the physical meaning of observables remains unclear in view of the ``no-hidden-variables'' theorems that tell us they cannot be thought of as physical quantities. The merit of Bohmian mechanics \citep{Bohm52,DT09} is to provide a full, clear, and satisfactory solution of these problems.

These problems persist in QED, not surprisingly, as NRQM should be contained in QED as a limiting case. On top of that, QED has problems that were absent in NRQM, as follows.

\section{Problems of QED}
\label{sec:problems}

\subsection{Ultraviolet Divergence} 

The Hamiltonian (and thus the unitary time evolution) is ill defined. The problem is milder if we only consider the time evolution from $t=-\infty$ to $t=+\infty$ (the $S$ matrix), but even that is ill defined; $S$ can heuristically be written as a series (a Dyson series), but the series does not converge. Even the terms in the series are ill defined to begin with, but there are proposals for how to change each term so as to make them well defined (renormalization and similar techniques \citep{Scha}); while some of these changes may appear ``ad hoc,'' other problems are worse than that.

\subsection{Born Rule for Photon Positions} 

NRQM provides, apart from a Hilbert space $L^2(\RRR^{3N})$ and a Hamiltonian, a Born rule $\rho(q)=|\psi(q)|^2$ for particle positions. In contrast, QED does not provide a Born rule for photon positions \citep[see][Sec.~7.3.9 for discussion]{Tum22} although experimentalists can measure photon positions. To be sure, we have a Born rule for photon quantum states that are planes waves or locally plane waves, and since quantum states tend to become locally plane waves as $t\to\pm\infty$, we can predict position probabilities for $t=\pm\infty$ but not in between. Some proposals in the literature \citep[such as][]{G57} may be good approximations but do not seem convincing as fundamental laws.

\subsection{Consequences of Excluded Negative Energy States} 

While the 1-particle Dirac Hamiltonian in $\Hilbert_1=L^2(\RRR^3,\CCC^4)$ has spectrum $(-\infty,-m]\cup [m,\infty)$ leading to a decomposition $\Hilbert_1=\Hilbert_{1+}\oplus \Hilbert_{1-}$ into the contribution of positive and that of negative energies, we usually regard only states without contributions of negative energy as physical. But $\psi\in\Hilbert_{1+}$ cannot be concentrated in a bounded region $A\subset \RRR^3$ \citep[Cor.~1.7]{Tha}, which leads to the question how $\psi$ collapses after the particle was detected in $A$. If the collapsed wave function is not strictly localized in $A$ (while detection probability density is still given by $|\psi|^2$), then the particle could be detected shortly afterwards far away from $A$, so the same particle could be detected in two spacelike separated regions, which sounds not believable. In particular, it apparently follows that either interaction locality (i.e., that the Hamiltonian contains no interaction terms between spacelike separated space-time points) is violated (so the detector could be triggered by a particle far away), or propagation locality (i.e., that wave functions propagate no faster than at the speed $c$ of light) is violated (so the particle's wave packet could travel from one detection region to the other), or both.

Variants of this problem are provided by the Hegerfeldt-Malament theorem \citep{Bor67,Heg74,Mal96,HC02,Beck23}, the Reeh-Schlieder theorem \citep{RS61,Beck23}, and a related problem that I described in \citep[Sec.~6 paragraph 4]{Tum21}. While some authors \citep[e.g.,][]{Sof11} think that propagation locality is only approximate and not strict, I find that hard to believe.

\subsection{Status of the Dirac Sea} 
\label{sec:sea}

\begin{figure}[ht]
\begin{center}
\includegraphics[width=.3 \textwidth]{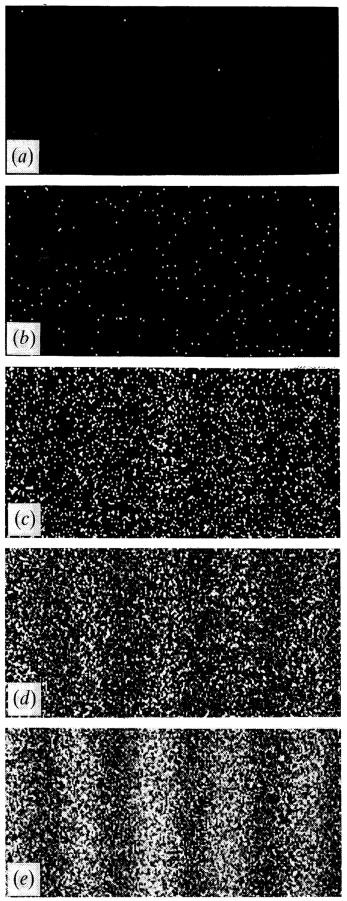}
\caption{A picture of actual results of a double-slit experiment of Tonomura et al., after (a) 10, (b) 100, (c) 3,000, (d) 20,000, (e) 70,000 electrons. (Reprinted from \citep{Ton89} with the permission of AIP Publishing.)}
\label{fig:tonomura}
\end{center}
\end{figure}

Everybody has an intuitive understanding of the Dirac sea, but for a physical theory we need more precision. Specifically, given that actual detectors can detect individual electrons (as shown in Figure~\ref{fig:tonomura} for a double-slit experiment) but do not detect any background of ``sea electrons,'' we need a Born rule for the positions of visible electrons (and positrons).

Standard presentations of QED do not provide such a Born rule. An obvious Born rule is to take, for any configuration $x\in\RRR^{3n}$ of $n$ electrons and any configuration $\overline{x}\in\RRR^{3\overline{n}}$ of $\overline{n}$ positrons,
\be\label{rhoobv}
\rho(x,\overline{x})=|\psi^{(n,\overline{n})}(x,\overline{x})|^2\,,
\ee
where $\psi$ lies in the standard Hilbert space of the quantum Dirac field,
\be\label{Hilbertdef}
\Hilbert=\Fock(\Hilbert_{1+})\otimes \Fock(C\Hilbert_{1-})
\ee
with the first factor representing electrons and the second positrons; here, 
\be
\Fock(\Hilbert)=\bigoplus_{n=0}^\infty \Hilbert^{\wedge n}
\ee
is the fermionic Fock space, $C$ the anti-unitary charge conjugation operator, and $C\Hilbert_{1-}$ equals in fact $\Hilbert_{1+}$. 
However, this obvious Born rule does not seem convincing \citep{Tum21}, also because of the Hegerfeldt-Malament theorem and related problems mentioned in the previous subsection. Alternative candidates are described in \citep{Tum21,Tum23} and in Section~\ref{sec:positron} below. 

Again, the problem is milder at $t=\pm\infty$, essentially because contributions to $\psi$ with different momenta tend to separate in space as $t\to\pm\infty$, so that the distribution of positions at $t=\pm\infty$ coincides with the momentum distribution, which is easier to define than the position distribution. For many experiments, including the one of Figure~\ref{fig:tonomura}, it is a good approximation to set $t=\infty$, but I am asking here for the correct position distribution at finite $t$: for the distribution to which the approximation is an approximation.

Some authors \citep[e.g.,][]{Mal96} say that all these problems indicate that the concept of particles does not make sense in QED. However, the problem here does not come from any theoretical definition or concept, it comes from the detection events. When I look at Figure~\ref{fig:tonomura}, I see individual, localized detection events as well as an obvious statistical distribution of these events, and I wonder which formula governs this distribution.

Moreover, if we lack a convincing Born rule for positions (or, equivalently, position operators), then we have a problem not just with detection probabilities, but also with how to link the quantum state to probabilities even for positions of macroscopic objects. For example, in a situation in which the needle of a measurement instrument can point either to the left or to the right, it is clear in NRQM how to translate the condition ``points to the left'' into a subset of the configuration space $\RRR^{3N}$ and thus into a subspace of the Hilbert space, a projection $P$, and a probability $\scp{\psi}{P|\psi}$. But in QED, if we do not have position operators to begin with, then on what grounds could we associate with the condition ``points to the left'' a subspace or operator or probability?

\subsection{The Shale-Stinespring-Ruijsenaars Problem} 

As a step towards defining the time evolution of the Dirac quantum field interacting with a quantized electromagnetic field, it is reasonable to first try to define the time evolution of the Dirac quantum field in the presence of an external, classical electromagnetic field $A_\mu$. As discovered by \citet{SS} \citep[see also][Sec.~10.3.3]{Tha}, the latter is well defined in the standard Hilbert space \eqref{Hilbertdef} if and only if the contribution from $A_\mu$ to the 1-particle Hamiltonian, $V:=\gamma^0\gamma^\mu A_\mu$, has off-diagonal part $V_{+-}:=P_{1+}VP_{1-}$ that is a Hilbert-Schmidt operator, $\tr(V_{+-}^\dagger V_{+-})<\infty$. This happens only, as shown by \citet{R77a,R77b}, when the magnetic field vanishes. Of course, that condition is very restrictive and generically violated. And if the time evolution is not well defined in an external $A_\mu$, there is little hope it could be well defined in a quantized $A_\mu$. 

One answer might be that \eqref{Hilbertdef} is not the right Hilbert space. After all, Fock space corresponds to a configuration space containing all configurations of \emph{finitely} many particles, and if configurations of \emph{infinitely} many particles occurred in nature, we might have to use corresponding Hilbert spaces instead of Fock space.

Alternatively, \citet{DDMS} pointed out that if we change the ``sea space'' (i.e., the subspace of $\Hilbert_1$ filled by the Dirac sea) by only finitely many dimensions, then the resulting Hilbert space as in \eqref{Hilbertdef} would, in a natural sense, stay the same; they were able to construct a Hilbert space for the quantum Dirac field if given, instead of a sea space, a suitable equivalence class of sea spaces. They went on to propose \citep{DDMS,DM16a} using different Hilbert spaces for different times or, more generally, for different Cauchy surfaces $\Sigma$ and showed that their choice of $\Hilbert_\Sigma$, which depends only on the magnetic field on $\Sigma$, leads to a unique (up to phase) time evolution of the quantum Dirac field, thus solving the Shale-Stinespring-Ruijsenaars problem. However, this approach does not lead to position operators or a Born rule for Dirac particles, as the obvious choice \eqref{rhoobv} would still require the choice of a sea space, for which there is no naturally selected candidate in the presence of an external $A_\mu$.

\subsection{Curved Space-Time}

In curved space-time $(\sM,g)$, the situation becomes worse. It becomes even more problematical than in Minkowski space-time to choose a subspace of $\Hilbert_{1\Sigma}$ as the sea space, as the Hamiltonian depends not only on the Cauchy surface $\Sigma$ but also on the lapse function, i.e., on how we push $\Sigma$ towards the future. A cousin of the Shale-Stinespring-Ruijsenaars problem returns, and I do not see how the strategy of \citet{DDMS} \citep[and][]{DM16a} could be applied in this case.

The problem is milder in asymptotically flat space-times, if we consider only very early and very late Cauchy surfaces. However, if we are after the fundamental laws of nature, then we also have to deal with times other than $\pm\infty$. Not to speak of the circumstance that our universe does not seem to be asymptotically flat, but singular in the past and asymptotically de Sitter in the future.

Likewise, the definition of position operators and the Born rule, at least for the analog of the ``obvious'' definition \eqref{rhoobv} in Minkowski space-time, depends on the choice of a sea space, so likewise it becomes even more problematical in curved space-time. Also in this context of quantum field theory in curved space-time, it is often said that the concept of particles breaks down, but again, the problem occurs regardless of whether particles are fundamental, it occurs as a problem concerning the detectors that we do have and, as pointed out at the end of Section~\ref{sec:sea}, concerning positions of macroscopic objects.

\section{And on Top of That}

Apart from the things about QED that downright fail to work, there are further oddities. 

\subsection{Complexity}

The most striking is perhaps that QED becomes extraordinarily and excessively complex when we try to clearly specify the laws of the theory. This is often hidden in a notation that specifies the Lagrangian in symbols such as ${A}_\mu(x)$ and ${\Psi}(x)$ for the field operators, but becomes painfully visible when we try to make explicit or define mathematically what exactly the theory says. For example, this is visible in \citep{DDMS} for just the problem of defining the time evolution of the quantum Dirac field in an external, classical $A_\mu$, or in the book of \citet{CTDRG97} that spends the majority of its pages on just defining the theory, without even aiming at mathematical rigor. 

This complexity is not only unpleasant, it is also odd as we usually expect that fundamental laws are simple. In contrast, I would readily expect that the computation of \emph{predictions} of quantities observed in experiment can be complicated; that is because the experiment itself may involve complicated details that need to be taken into account, but more importantly that computing the prediction may involve \emph{solving} fundamental equations of the theory, such as time evolution equations. But that is different from \emph{stating} the fundamental equations, which should be simple. I suspect that the positivistic attitude of the orthodox view of quantum physics contributes to this difficulty: for a realist, there is a big and clear difference between stating the laws that govern the fundamental physical reality and making predictions for experiments; for a positivist, in contrast, the distinction is blurred, and making predictions is the only activity worthy of a scientist. So, what realists like me would like to see is a set of simple fundamental laws that could be regarded as the definition of the theory and as the ingredients from which to derive empirical predictions.

\subsection{Wave Functions}

Another confusing aspect is that it is common in QED to talk about field operators, but not about wave functions. Of course, this is in a way a matter of representation, as one could say that wave functions are included in the field operator picture because the field operators will act on some Hilbert space $\Hilbert$, and vectors in $\Hilbert$ should be the analog of wave functions (or positive functionals on operator algebras the analogs of density matrices). And yet, something is odd. Wave functions or state vectors are central elements of quantum physics, and some arguments such as the Pusey-Barrett-Rudolph theorem \citep{PBR} support that they exist in reality. For comparison, when, in courses on classical physics, the electromagnetic field gets introduced, we pay much attention to how to represent it in a covariant way and to what kind of tensor field it is, such as $F_{\mu\nu}=-F_{\nu\mu}$ and the consequences. But in QED, all attention goes to the field operators, none to the wave function. I will outline in Section~\ref{sec:LP} a proposal for how to represent the wave function of QED in a covariant way.

\subsection{Fields or Particles}

Let us come back to the problems that QED shares with NRQM and that circle around the lack of a clear ontology. The fact that Bohmian mechanics solves the problems for NRQM is why we would like to do something similar for QED. A basic question we face is whether to choose a field ontology or particle ontology. Setting up a particle ontology requires a Born distribution (such as given by position operators) over particle configurations, which leads to the problems mentioned above; but the distributions we see in detections such as Figure~\ref{fig:tonomura} suggest that these problems are solvable. Bohm himself proposed already in 1952 \citep{Bohm52} to use a particle ontology for fermions and a field ontology for bosons. In my humble opinion, this proposal is not entirely convincing \citep[see also][Sec.~6.5.2 for discussion]{Tum22}. The asymmetry in treating fermions and bosons does not seem justified from the difference between permutation anti-symmetry and permutation symmetry. Then, the field ontology itself seems unnatural in view of boson detection events. Next, it is not clear how to formulate a Born rule for field configurations because the space of field configurations is $\infty$-dimensional, there are problems with obtaining a notion of volume in $\infty$-dimensional space, but a notion of volume is needed when we want to take the function $|\psi|^2$ to be a probability \emph{density}. For fermions, it is unclear how a field ontology could even be set up, as the field operators cannot simply be multiplication by field variables since fermionic field operators anti-commute whereas multiplication operators commute. Furthermore, it is unclear for fermions which macro-configurations would correspond to which field configurations. In view of all of these reasons together, I am more inclined to expect that the final Bohmian version of QED will use a particle ontology for both the fermions and the bosons.

\section{Particle Position Picture, Starting from the Work of Landau and Peierls}
\label{sec:LP}

Let us turn to some considerations about what a Bohm-style QED could look like. Lev Landau and Rudolf Peierls published a remarkable paper \citep{LP30} in which they proposed a formulation of QED in the particle-position representation for both electrons and photons.  I think that if this approach can be carried to the end, then this representation is a particularly natural way of framing QED. Landau and Peierls themselves did not follow up later on this approach, perhaps in part because they did not have a Born rule for photons, and perhaps also because their equations were not manifestly Lorentz or gauge invariant. However, they become invariant \citep{LT23} when formulated in terms of multi-time wave functions \citep{LPT:2020}, which here are of the form
\be\label{Psiform}
\Psi=\Psi^{(m,n)}_{s_1...s_m,\mu_1...\mu_n}(x_1...x_m,y_1...y_n)\,,
\ee
where $m$ is the number of electrons (positrons are left out of the model), $n$ is the number of photons, both $m$ and $n$ are variable as in Fock space, $x_j\in\RRR^4$ is a space-time point considered as the location of the $j$-th electron, $y_k\in\RRR^4$ a space-time point for the $k$-th photon, $x_j$ and $y_k$ are required to be mutually spacelike separated but otherwise arbitrary, $s_j\in\{1,2,3,4\}$ is the index (associated with the $j$-th electron) for the 4 components of a Dirac spinor, and $\mu_k\in\{0,1,2,3\}$ the index (associated with the $k$-th photon) for the 4 components of a space-time vector as in $A_\mu$. The wave function of a single photon is taken to satisfy the complexified Maxwell equation, up to particle creation and annihilation terms. In fact, the appropriate multi-time evolution equations, one for each particle, read \citep{LT23}:
\begin{subequations}\label{LP}
\be\label{LPx}
(i\gamma^\mu_j\partial_{x_j,\mu}-m_x)\Psi^{(m,n)}(x_1...x_m,y_1...y_n)
= e_x\sqrt{n+1} \:\gamma^\rho_j\: \Psi^{(m,n+1)}_{\mu_{n+1}=\rho}(x_1...x_m,y_1...y_n,x_j)\,,
\ee
\begin{multline}\label{LPy}
2\partial_{y_k}^\mu\partial^{~}_{y_k,[\mu}\Psi^{(m,n)}_{\mu_k=\nu]}(x_1...x_m,y_1...y_n) =\\
\frac{e_x}{\sqrt{n}}\sum_{j=1}^m \delta^3_{\mu}(y_k-x_j) \: \gamma_j^\mu \gamma_{j\nu} \:\Psi^{(m,n-1)}_{\widehat{\mu_k}}(x_1...x_m,y_1...y_{k-1},y_{k+1}...y_n)\,.
\end{multline}
\end{subequations}
Here, $m_x$ and $-e_x$ are the mass and charge of the electron, $\gamma_j^\mu$ means $\gamma^\mu$ acting on the index $s_j$, $[\mu\nu]$ anti-symmetrization in the index pair as in $S_{[\mu\nu]}=\tfrac12(S_{\mu\nu}-S_{\nu\mu})$, $\delta^3_\mu$ the Lorentz-covariant 3d delta function \citep{LT23}, and $\widehat{\mu_k}$ that the index $\mu_k$ is omitted. The first equation \eqref{LPx} is the Dirac equation with an additional term on the right-hand side resembling the electromagnetic potential $A_\mu$ acting on $\Psi$, but now provided by the wave function of the next photon, which is part of $\Psi$ itself. The second equation \eqref{LPy} is the Maxwell equation for $A_\mu(y)$ given by $\Psi$ as a function of $y=y_k$ with index $\mu=\mu_k$, and with an additional term on the right-hand side representing the emission of photon $k$ by electron $j$. Although the notation that I have used in order to make all variables and indices visible leads to long expressions, these equations are really of a remarkable simplicity. Actually, their degree of simplicity would seem quite appropriate for fundamental laws of nature. (But keep in mind that nothing has been done here to deal with the ultraviolet divergence or to incorporate pair creation or positrons at all.) This form of the wave function and the equations appears so natural that I am inclined to think that this $\Psi$ must be what quantum states really are in nature.

\section{Positrons}
\label{sec:positron}

Landau and Peierls did not have in mind Bohmian trajectories, but such trajectories can be introduced in standard ways \citep{Tum22} once we know how to compute the probability current 4-vector field, that is, once we have the full space-time version of the Born rule. Since we do not have that for photons, I will leave them aside and turn to Dirac particles. For a single Dirac particle, the current is given by
\be
j^\mu= \overline{\psi} \gamma^\mu \psi\,.
\ee

The question arises how to treat positrons. One possibility that has been considered by several authors \citep{BH,CS07,DEO} is to take the Dirac sea literally and introduce an infinitude of Bohmian particles as electrons of negative energy (or possibly introduce a cut-off that will make the dimension of $\Hilbert_{1-}$ finite). It would be of interest to study whether a Bohmian theory of an actual infinitude of particles can work mathematically. Even heuristically, it is not very clear how to think of such a particle configuration. Is it a countably infinite set of points in $\RRR^3$? Why not a set whose complement is countably infinite? Some heuristics \citep[Sec.~5]{Tum21} suggest that the Dirac sea should have exactly two particles at every space point $\vx\in\RRR^3$ except for a countable set of $\vx$'s that each host 0, 1, 3, or 4 particles. The character of sea configurations is also relevant to the question whether it would somehow be visible from the configuration at which places there is a particle missing or a particle in excess, and thus which places should light up in a detector image such as Figure~\ref{fig:tonomura}.

Let me turn to two alternative candidates for treating positrons, corresponding to alternatives for the Born rule and for the position operators.

\subsection{A New Possibility}

The first \citep{Tum21} is based on defining, for every region $A\subset \RRR^3$ in physical 3-space, a charge operator $Q(A)$ by
\be
Q(A)=-e_x \int_A d^3\vx \sum_{s=1}^4 :\Psi_s^\dagger(\vx) \, \Psi_s(\vx):~~,
\ee
where $-e_x$ is again the electron charge, $\Psi_s(\vx)$ is the Dirac field operator associated with location $\vx$ and spin component $s$, and colons mean normal ordering. According to heuristic reasoning, $Q(A)$ has spectrum in the set $e_x\ZZZ$ of integer multiples of $e_x$, and the canonical anti-commutation relations for $\Psi_s(\vx)$ imply that $Q(A)$ and $Q(A')$ commute for any $A$ and $A'$, so all of these charge operators can be diagonalized simultaneously. The joint spectrum should then, at least when $\RRR^3$ is replaced by a box of finite volume, consist of signed configurations, i.e., configurations of finitely many points, of which some are occupied by a positive charge and some by a negative charge. Since the ``sea state,'' i.e., the vector $\Omega=|0\rangle \otimes |0\rangle$ in \eqref{Hilbertdef} usually called the ``vacuum state,'' has $\scp{\Omega}{Q(A)|\Omega}=0$ for all $A$ but is not an eigenvector of $Q(A)$, the corresponding Born distribution in this picture is actually not concentrated on the empty configuration; rather, in this picture $|\Omega\rangle$ corresponds to a random number of electron and positron particles in random places. 

However, an attempt to rigorously define the operators $Q(A)$ has failed \citep{CT23}, so the viability of this approach is dubious.

\subsection{Another New Possibility}

A second proposal \citep{Tum23} can be summarized very briefly by saying that, contrary to what Dirac proposed, \emph{a positron is a Dirac particle with negative energy, rather than a hole among them.} Some remarks:

\begin{itemize}
\item We may note first that according to the 1-particle Dirac equation, a wave packet consisting purely of negative energy contributions will move in an external electromagnetic field exactly like a wave packet of purely positive contributions but with the opposite charge. That is because the charge conjugation operator maps negative energy contributions to positive ones and vice versa, and maps the Dirac Hamiltonian to the one with opposite charge. 

\item An objection that comes to mind immediately is that if particles of negative kinetic energy existed, so that the free Hamiltonian was unbounded from below, a particle could repeatedly emit radiation with positive energy at the expense of pushing its own energy toward $-\infty$. Then again, if negative energy is allowed, the particle could also emit radiation of negative energy and thereby push its own energy toward $+\infty$, which suggests that, as in a random walk, the actual energy might not grow quickly to either $+\infty$ or $-\infty$. Although at first, this situation may appear rather unstable, it looks like it would take an infinite time before an infinite amount of energy gets radiated off; on top of that, there is an argument \citep{Tum23}, based on a mapping (that I will call $W$) translating the negative energy states into holes in the Dirac sea, to the effect that the situation is more stable than it may seem, in fact no less stable than the usual Dirac sea picture.

\item For a Hamiltonian $H$ not bounded from below, there exists no normalized thermal equilibrium density matrix
\be\label{rhocan}
\rho = \frac{1}{Z}e^{-\beta H}
\ee
for any inverse temperature $\beta>0$. On the other hand, in QED we are after a theory of the universe (or a hypothetical universe without non-electromagnetic interactions), and we would not consider the whole universe in thermal equilibrium anyway. In addition, since the mentioned mapping $W$ translates into the Dirac sea picture, which possesses density matrices such as \eqref{rhocan}, there presumably remains some type of universal thermal equilibrium state.

\item Another objection that quickly comes to mind is that an external electromagnetic field may lift a negative energy state to positive energy. In the Dirac sea picture, that sounds just right: it creates a hole among the negative energy states, along with a particle of positive energy; in other words, it creates an electron-positron pair. In a model in which positrons are just Dirac particles with negative energy, this process would seem to entail that a positron has disappeared and an electron has appeared, in violation of charge conservation. However, that changes if we make the additional hypothesis that \emph{in the time evolution equations, there are explicit terms for the creation and annihilation of electron-positron pairs.} That is, two further terms are introduced in \eqref{LP}, coupling $\Psi^{(m,n)}$ to $\Psi^{(m+2,n-1)}$ and $\Psi^{(m-2,n+1)}$. 

\item Two nice traits of the proposal are that it ensures interaction locality and propagation locality (circumventing the Hegerfeldt-Malament theorem because $H$ is not bounded from below), and that it extends in a straightforward way to curved space-time, even if not asymptotically flat. If positrons are just Dirac particles of negative energy, and if the concept of positive or negative energy becomes ill-defined in some cases, then in those cases there would not be a fundamental fact about whether a given particle is an electron or a positron---it is just a Dirac particle. Which particles we call electrons and which positrons may thus be a categorization that is meaningful and useful in flat but not in curved space-time, or only at $t=\pm\infty$ but not in between. That is, according to this possibility, it is fundamentally not defined whether a given Dirac particle is an electron or a positron, and only in certain regimes can particles be consistently labeled as electrons or positrons. (This situation is similar to the possibility explored in \citep{GTTZ05} that particles of different species are no different in the fundamental ontology.)
\end{itemize}
The question whether such a model can be empirically adequate requires further analysis.

\section{Synthesis}

The Bohm-type theory of QED that I am aiming at has particle trajectories for fermions and photons, although satisfactory equations for photon trajectories have not been identified yet. It uses the Landau-Peierls formulation of QED in the particle position representation in the form \eqref{LP} for a multi-time wave function \eqref{Psiform}, with further terms governing pair creation and annihilation. The Bohm-like trajectories depend on a preferred foliation of space-time into spacelike hypersurfaces \citep{HBD}, while the wave function does not. The creation of Bohmian particles corresponds to stochastic jumps in the configuration \citep{DGTTZ}; the jump rates are determined by the probability current. The possibility to treat positrons as Dirac particles of negative energy seems worth considering. 
The problem of ultraviolet divergence might affect all of the theories considered here in the same way; then again, interior-boundary conditions \citep{TT15} have ameliorated this problem in other models and might also help here.

I expect that the empirical predictions of the theory I have outlined coincide with what is usually regarded as the predictions of QED; after all, both are essentially based on the same time evolution expressed in different ways, so that the same series expansions etc.\ are applicable. Still, the theory I have outlined would provide a clear ontology, it would seem much simpler than previous attempts at a fundamental formulation of QED and at the same time better suited to provide an evolution between finite times, also in curved space-time. That would be a satisfactory formulation of QED.

\bigskip

\noindent{\it Acknowledgments.} I thank Stefan Teufel for helpful discussions.

\end{document}